\documentclass[universe,article,accept,pdftex,moreauthors]{Definitions/mdpi} 
\firstpage{1} 
\pubvolume{1}
\issuenum{1}
\articlenumber{0}
\pubyear{2022}
\copyrightyear{2022}
\externaleditor{\hl{Academic Editor: Lorenzo Iorio} 
} 
\datereceived{08 September 2022} 
\dateaccepted{16 October 2022} 
\datepublished{} 
\hreflink{https://doi.org/} 

\Title{Launching the VASCO Citizen Science Project}

\TitleCitation{Launching the VASCO Citizen Science Project}

\Author{Beatriz Villarroel $^{1,2,}$, Kristiaan Pelckmans $^{3}$, Enrique Solano $^{4}$, Mikael Laaksoharju $^{3}$, Abel Souza $^{5}$, Onyeuwaoma Nnaemeka Dom $^{6}$, Khaoula Laggoune $^{7}$, Jamal Mimouni $^{8}$, Hichem Guergouri $^{9}$, Lars Mattsson $^{1}$, Aurora Lago Garc\'ia $^{10}$, Johan Soodla $^{3}$, Diego Castillo $^{3}$, Matthew E. Shultz $^{11}$, Rubby Aworka $^{12,13}$, S\'ebastien Comer\'on $^{2,14}$, Stefan Geier $^{2,15}$, Geoffrey W. Marcy $^{16}$, Alok C. Gupta $^{17}$, Josefine Bergstedt $^{18}$, Rudolf E. B\"{a}r $^{19}$, Bart Buelens $^{20}$, Emilio Enriquez $^{21}$, Christopher K. Mellon $^{22}$, Almudena Prieto $^{2,14}$, Dismas Simiyu Wamalwa $^{23}$, Rafael S. de Souza $^{24}$ and Martin J. Ward $^{25}$}

\AuthorNames{Beatriz Villarroel, Kristiaan Pelckmans, Enrique Solano, Mikael Laaksoharju, Abel Souza, Onyeuwaoma Nnaemeka Dom, Khaoula Laggoune, Jamal Mimouni, Hichem Guergouri, Lars Mattsson, Aurora Lago Garc\'ia, Johan Soodla, Diego Castillo, Matthew E. Shultz, Rubby Aworka, S\'ebastien Comer\'on, Stefan Geier, Geoffrey Marcy, Alok C. Gupta, Josefine Bergstedt, Rudolf E. B\"{a}r, Bart Buelens, J. Emilio Enriquez, Christopher K. Mellon, M. Almudena Prieto, Dismas Simiyu Wamalwa, Rafael S. de Souza and Martin J. Ward}

\AuthorCitation{Villarroel, B.; Pelckmans, K.; Solano, E.; Laaksoharju, M.; Souza, A.; Dom, O.N.; Laggoune, K.; Mimouni, J.; Guergouri, H.; Mattsson, L.; et al.}

\address{%
$^{1}$ \quad Nordita, KTH Royal Institute of Technology and Stockholm University, Roslagstullsbacken 23, SE-106 91 Stockholm, Sweden; lars.mattsson@su.se 
\\
$^{2}$ \quad Instituto de Astrof\'isica de Canarias, Avda V\'ia L\'actea s/n, E-38205 La Laguna, Tenerife, Spain; lsebasti@ull.edu.es (S.C.); stefan.geier@gtc.iac.es (S.G.); aprieto@iac.es (M.A.P.)\\

$^{3}$ \quad Department of Information Technology, Uppsala University, Box 337, SE-751 05, Uppsala, 
 Sweden; kristiaan.pelckmans@physics.uu.se (K.P.); mikael.laaksoharju@it.uu.se (M.L.); johan.soodla@gmail.com (J.S.); diegocasmo@gmail.com (D.C.)\\

$^{4}$ \quad Departamento de Astrof\'isica, Centro de Astrobiolog\'ia (CSIC/INTA), P.O. Box 78, 
E-28691 Villanueva de la Ca\~{n}ada, Spain; esm@cab.inta-csic.es\\

$^{5}$ \quad Autonomous Distributed Systems Lab, Department of Computing Science, Umeå Universitet; 901 87 Umeå, Sweden; abel@cs.umu.se\\

$^{6}$ \quad Center for Basic Space Science, National Space Research and Development Agency, Post code: 410002
or P.O. Box 2022, Nsukka, Enugu
, Nigeria; emekadonn@gmail.com\\

$^{7}$ \quad Sirius Astronomy Association, Sirius Astronomy Association, BP 18, Cité du 20 Aout, Constantine, 25000, Algeria
; khaoula-laggoune@hotmail.com\\

$^{8}$ \quad Department of Physics, University of Constantine 1, LPMPS \& CERIST,  25000,  
Constantine, 
 Algeria; jamalmimouni@yahoo.com\\

$^{9}$\quad Science of Matter Division, Research Unit in Scientific Mediation, 
 CERIST, 25016, Constantine, Algeria; guergouri.hichem.92@gmail.com (H.G.); paurora.lago.fisicaquimica@iestartessos.net (A.L.G.)\\

$^{10}$\quad 
 IES Tartessos, Barriada Hiconsa s/n 41900, Camas, Sevilla, Spain\\  

$^{11}$\quad Department of Physics \& Astronomy, University of Delaware, 217 Sharp Lab Newark, DE 19716, USA
; matt.shultz@gmail.com\\

$^{12}$\quad Ghana Space Science and Technology Institute, Atomic-Haatso Road, Kwabenya, Box LG 80 233 Accra, Ghana 
; aworka123@gmail.com\\

$^{13}$\quad African Institute for Mathematical Sciences Ghana, Accra (AIMS Ghana), University of Ghana, Summerhill Estates Ltd, Rd, Santeo, P.O. Box LG 25 Legon, Accra, Ghana\\

$^{14}$\quad Departamento de Astrofísica, Universidad de La Laguna, \mbox{E-38206 La Laguna, Tenerife, Spain}\\

$^{15}$\quad Gran Telescopio Canarias (GRANTECAN), Cuesta de San José s/n, \mbox{38712 Breña Baja, La Palma, Spain}\\

$^{16}$\quad Space Laser Awareness, 3883 Petaluma Hill Rd, Santa Rosa, CA 95404 USA
; geoff.bnb@gmail.com\\

$^{17}$\quad Aryabhatta Research Institute of Observational Sciences (ARIES), Manora Peak, Nainital 263 001, India; acgupta30@gmail.com\\

$^{18}$\quad Freelance, 752 17 Uppsala, Sweden 
 Sweden; josefine\_bergstedt@hotmail.com\\

$^{19}$\quad Institute for Particle Physics and Astrophysics, ETH Zürich, Wolfgang-Pauli-Strasse 27, \mbox{CH-8093 Zürich, Switzerland}; rbaer25@gmail.com\\

$^{20}$\quad Flemish Institute for Technological Research (VITO), 2400 Mol, Belgium; bbuelens@gmail.com\\

$^{21}$\quad Department of Astrophysics/IMAPP, Radboud University Nijmegen, Nijmegen,  6525 AJ, The Netherlands 
; jeenriquez@gmail.com\\

$^{22}$\quad Galileo project affiliate. The \\ 
; christophermellon1@icloud.com\\

$^{23}$\quad Meru 
Physics Department, University of Science and Technology, Meru P.O. Box 972-60200, Kenya; dismasw@yahoo.com\\

$^{24}$\quad Key Laboratory for Research in Galaxies and Cosmology, Shanghai Astronomical Observatory, Chinese Academy of Sciences, 80 Nandan Rd., Shanghai 200030, China; rafael.2706@gmail.com\\

$^{25}$\quad Centre for Extragalactic Astronomy, Department of Physics, Durham University, South Road, \mbox{Durham DH1 3LE, UK}; martin.ward@durham.ac.uk\\

}

\corres{Correspondence: beatriz.villarroel@su.se 
}

\abstract{The Vanishing \& Appearing Sources during a Century of Observations (VASCO) project investigates astronomical surveys spanning a time interval of 70 years, searching for unusual and exotic transients. We present herein the VASCO Citizen Science Project, which can identify unusual candidates driven by three different approaches: hypothesis, exploratory, and machine learning, which is particularly useful for SETI searches. To address the big data challenge, VASCO combines three methods: the Virtual Observatory, user-aided machine learning, and visual inspection through citizen science. Here we demonstrate the citizen science project and its improved candidate selection process, and we give a progress report. We also present the VASCO citizen science network led by amateur astronomy associations mainly located in Algeria, Cameroon, and Nigeria. At the moment of writing, the citizen science project has carefully examined 15,593 candidate image pairs in the data (ca. 10\% of the candidates), and has so far identified 798 objects classified as ``vanished''. The most interesting candidates will be followed up with optical and infrared imaging, together with the observations by the most potent radio telescopes.}

\keyword{surveys; transients; SETI; citizen science 
} 

\begin{document}
  

\section{Introduction}

Anomalous objects in astronomy are a gold mine for expanding our knowledge about extreme physical conditions and identifying new astrophysical phenomena. Anomalies have always fascinated astronomers and many important discoveries were first regarded as such. For~instance, when the first optical spectra of the radio-emitting quasars 3C 273 and 3C 48 were acquired, astronomers encountered weird and unusual spectra that they considered anomalous, only to soon understand these quasi-stellar objects were in fact highly redshifted~\cite{Sandage1963,Schmidt1963}. Likewise, when the first pulsars were discovered~\cite{Hewish1968}, the~unexpected pulsating radio signals were considered so unlikely that Little Green Men were suggested as a serious possibility. With~further investigation, astronomers ultimately developed an understanding of the physics underlying these entirely natural, albeit extreme~objects. 

Some anomalies have come to stay with us as interesting examples of rare astrophysical objects. An~example is {\it Przybylski’s star} \cite{Przybylski1963}, a~variable star showing unusual amounts of iron and nickel in its spectrum while having high abundances of, e.g.,~strontium and uranium. Another is the well-known transient {\it $\eta$ Carinae}, whose lightcurve showed a giant outburst followed by a slow fading over decades. Other previously well-known astrophysical anomalies have fallen from prominence, following explanation of the underlying physics or identification of the supposed anomaly as an artifact---for example {\it Halton Arp's redshift anomalies} \cite{Arp1987,Burbidge2001} now believed to be chance overlaps in images, but~which were once the subject of a grand quarrel among cosmologists in the~1980s. 

Some recent anomalies have received much attention in the media; for example {\it `Oumuamua}, a~cigar-shaped interstellar visitor that followed a non-gravitationally bound orbit and does not seem like the most common comet~\cite{Meech2017}, or, {\it Tabby's Star} \cite{Tabby2016}, a~star with an unusual slow dimming caused by obscuration due to an uneven ring of surrounding dust~\cite{Meng2017}, and {\it Ross 128}, a~red dwarf, also figured in the media due to its unusual emission. These examples may need another few years of examination before we understand the key details of the physical mechanisms involved, and~it is possible that once we do understand them, we will no longer even consider them anomalies. The~same goes for {\it Fast Radio Bursts} (FRBs), a~completely novel class of poorly understood transients, for~which the responsible mechanism(s) remain a hotly debated topic. Already in the early 2000s, the~importance of a state-of-the-art development of methods to identify fascinating anomalies was discussed by, e.g.,~Djorgovski~et~al. (2000, 2001) \cite{Djorgovski2000,Djorgovski2001}. The~importance of anomalies with respect to Searches of Extra Terrestrial Intelligence (SETI) was carefully discussed in the same papers. A~recent work that compiles a list of anomalies is the {\it Breakthrough Listen Exotica Catalog}~\cite{Lacki2020}.

One of the successful ways of identifying anomalies is through citizen science projects, where volunteers help scientists in scrutinizing the extremely large datasets assembled by astrononomical surveys. Citizen science projects have already earned a good reputation by leading to interesting discoveries. We can thank the {\it Galaxy Zoo} project~\cite{Lintott2008,Lintott2011} for improving our understanding of galaxy evolution, utilizing visual inspection of images of galaxies acquired by the Sloan Digital Sky Survey (SDSS) and subsequent classification according to the most suitable morphological class. An~important consequence of this citizen science project was the discovery of ``Green peas'', a~rare class of galaxies with very low masses and high star-formation rates that looked round and green~\cite{Cardamone2009}. Interesting astrophysical anomalies such as, e.g.,~{\it Hanny's Voorwerp}~\cite{Jozsa2009}---a rare quasar ionization echo---and {\it Tabby's Star} (KIC 8462852) are the results of such citizen science searches. Citizen science projects are now getting competition from machine learning-based identification of anomalies~\cite{Baron2017,Giles2019}. Machine learning is certainly helpful in the analysis of giant datasets, but~while modern computing and automated routines can aid the identification of unusual objects, it cannot yet replace the human pattern recognition competence honed by millions of years of evolution. 

The most famous citizen science project, the~Galaxy Zoo, allowed for an entire community of citizen science projects to be assembled. The~Zooniverse has inspired millions of users to join scientists of different fields in the exploration of nature. The~current Zooniverse projects operate in different ways where a registered (or even an unregistered) user can contribute. The~total number of classifications exceeds half a billion. Current projects within time domain astronomy give the user different roles. {\it Backyard worlds} shows blinking images taken at different times that permits a user to identify fast-moving objects in WISE/NEOWISE data and mark them in the data. {\it Planet Hunter's Transiting Exoplanet} Survey Satellite (TESS) survey lets the user identify and mark possible transits. {\it Superwasp variable stars} asks the user to classify light curves into either a well-known category or mark it as junk. {\it Supernova Hunters} uses a target image, an~older image from Pan-STARRS, and~the resulting difference images, and~asks the user to distinguish between real supernovae and bogus~detections.

Herein, we present the citizen science project related to the Vanishing \& Appearing Sources during a Century of Observations (VASCO) project 
 \endnote{\url{http://iactalks.iac.es/talks/view/1358}}$^{,}$\endnote{\url{http://vasconsite.wordpress.com  Access on October 22 2022.}} \cite{Villarroel2016,Villarroel2020}. VASCO is a research program that compares historical data from 1950s sky catalogs to modern sky surveys. Using a 70-year temporal baseline, we target stars that may have appeared or vanished during the last seven decades---extreme phenomena that may be so rare that they are missed by transient sky surveys due to the short time windows. Simultaneously, more conventional strong one-epoch transients may be identified with the same approach. \mbox{Villarroel~et~al. (2020) \cite{Villarroel2020}} identified 150,000 candidate objects that need to be visually inspected, based on the cross-match methods described by Soodla (2019) \cite{Soodla2019}. Of~these, we inspected about 15\% with the help of images from the Sloan Digital Sky Survey (SDSS). We found about $\sim$100 red point sources where nearly all were visible in only one epoch and in the red images of the POSS-I survey. The~shapes and time scales involved rule out solar system objects, variable stars, low-redshift supernova, and~AGN. 

It is remarkable that the identified point sources so far have no counterparts in modern transient surveys such as the intermediate Palomar Transient Facility (iPTF), the~Gaia survey, or~the Catalina Sky Survey. These surveys tend to observe hundreds of short transients only visible in one image in one night, using well-calibrated and homogeneous CCD data. These automated surveys tend to discover thousands of flaring and erupting stars, cataclysmic variables, supernovae, GRB afterglows, variable or erupting active galactic nuclei (AGN), and~microlensing events. The~transients detected by photometry that are deemed interesting enough to follow up on usually also get a spectrum taken to determine its nature. In~this way, tens of thousands of transients have been discovered and~categorized.

One may expect that if the red transients were caused by  variable or flaring stars, the~occurrence would happen once more within the time window of the transient surveys. The~fact that these point sources have escaped all transient surveys so far suggests they were not  a phenomena with repetition time scales less than a few years. For~example, a~star that flares up once per week would be found by these transient surveys. This suggests that our red transients are not among the most common among transient phenomena, or~at least are not among the type of transients the big surveys are interested in following up on.

In the VASCO citizen science project, we use images from different sky surveys to search for both vanishing and appearing stars, as~well as transients. Some of the objects found may be similar to what is found automatically by the large transient surveys, and~some will be~different. 

The VASCO citizen science project combines three different strategies to search for anomalous objects:

\begin{enumerate}
    \item Direct identification: did the star vanish or appear? The simplest of all questions is approached here with a classical citizen science approach. This part of the citizen science project follows the concept of {\it hypothesis-driven science};
    \item AI-based selection of the most interesting candidates. With~the help of the users whose mission is to match two images taken at two different times, the~underlying AI is learning to identify the most interesting images in the large dataset. The~machine-learning algorithm is described by Pelckmans~et~al. (in prep); 
 \cite{Pelckmans2021}.
    \item User-driven exploration: here the user is allowed to decide what he or she defines as ``interesting'' in an {\it exploratory approach to science}. The~user is provided with 10 different images of the same field in different bands and the user is able to comment on anything of interest. This allows for experts with different backgrounds to examine the image data from the perspective of their own interests, and~highlight anything worth closer examination.
\end{enumerate}

\begin{figure}[H]
  \includegraphics[scale=1]{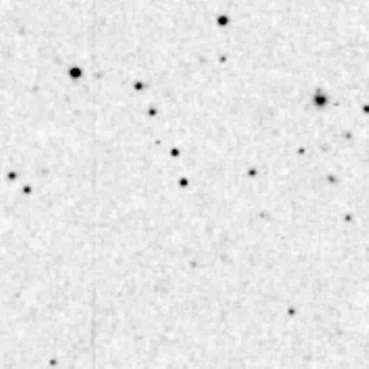}
  \includegraphics[scale=1]{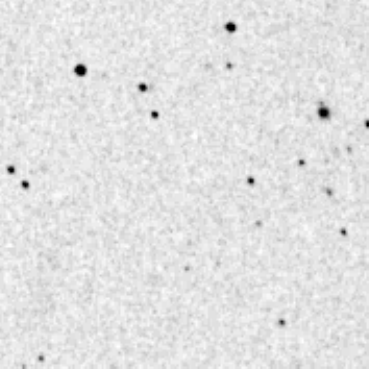}
  \caption{\label{simult} An example of an anomalous observation found by VASCO. Two sources disappear within a short time. The~left POSS-I E (red) image was taken in the early spring of 1950. The~right POSS-I E (red) image was taken six days~later.}
   \end{figure}

One of the main topics on the {\it Technoclimes}~\endnote{\url{https://technoclimes.org/}} workshop 2020 was indeed hypothesis-driven vs. exploratory-driven searches for anomalies. Papers by \mbox{Sheikh~et~al. (2020) \cite{Sheikh2020}} and Singam~et~al. (2020) \cite{Singam} discuss hypothesis-driven vs.\ exploratory-driven SETI. An~example of an anomalous observation detected in VASCO can be seen in Figure~\ref{simult}.

In this paper, we first describe the structure of the VASCO Citizen Science Project and the philosophy of its data analysis methodology. We then present the VASCO web interface~\endnote{\url{http://ml-blink.org}} \cite{Pelckmans2021} and its use. Finally, we present the VASCO citizen science network, which aims to make the citizen science project both pedagogically fruitful and entertaining for students and amateur astronomers of all ages and levels of knowledge, whilst generating data analysis products of use to the scientific community. 

\section{Image~Preprocessing}
\label{sec:data}

The web interface is currently using the original sample from 150,000 candidates presented by Villarroel~et~al. (2020) \cite{Villarroel2020}, obtained through a $30''$ cross-matching of the USNO and Pan-STARRS catalogs~\cite{Soodla2019}. This sample of 150,000 candidates may host even more fascinating objects than the transients already found, and~maybe even one of the real vanishing objects we are looking for. However, this candidate list is large and in need of~preprocessing.

\textls[-25]{As is true for any archive-based project, VASCO can be affected by the following~problems:}
\begin{itemize}
    \item Discovery: Where does the information of interest reside?
    \item Access: Each astronomical archive has developed its own data access system which makes data querying quite cumbersome if the number of services to be consulted \mbox{is high.}
    \item Representation: Most of the time, data gathered from different archives cannot be directly compared. In~the case of images, different sky coverage, orientation and/or pixel size demand a pre-processing analysis for the comparison to be possible. 
\end{itemize}

\begin{figure}[H]
   \resizebox{\hsize}{!}{
   \includegraphics[scale=0.65]{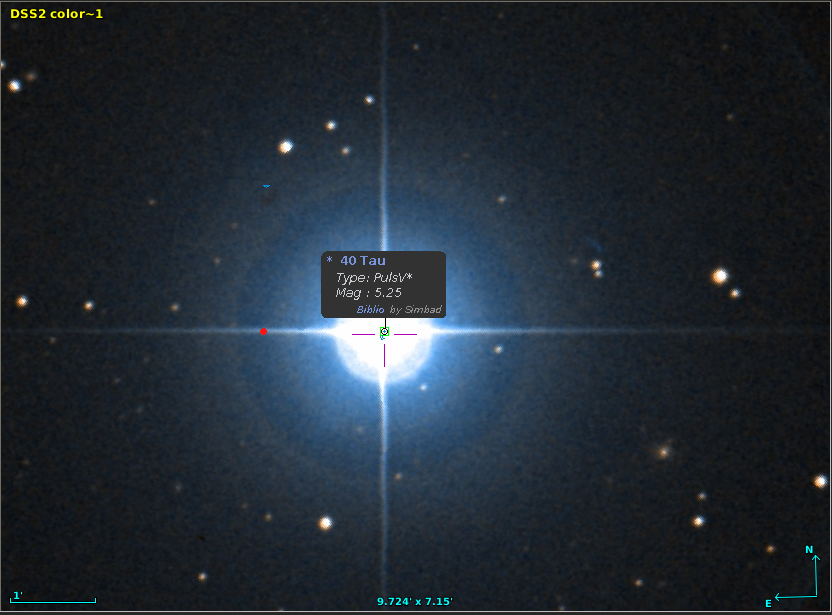}}
  \caption{\label{spikes} Spurious entry in the sample of 150,000 candidates lying on the spikes of a bright star (red dot). These type of sources are removed during the preprocessing~phase.}
   \end{figure}

All these issues can be largely alleviated if a Virtual Observatory methodology is considered. The~Virtual Observatory~\endnote{\url{http://www.ivoa.net}} (VO) is an international initiative that was born in the year 2000 and whose main goal is to guarantee easy and efficient access and analysis of the information hosted in astronomical archives. In~particular we have taken advantage of VO to provide citizens with as clean a sample of objects as possible, where most of the instrumental artifacts have been filtered out. In~this context, two actions were accomplished:

\begin{itemize}
    \item Removal of USNO sources lying on the vicinity of bright stars: Diffraction spikes are lines radiating from the centers of bright astronomical sources. These features are generated when the incoming light is diffracted by the structure which upholds the secondary mirror in reflecting telescopes. 
    To identify and remove the sources of the sample of 150,000 candidates associated with diffraction spikes, we visually inspected several hundred images to find the distribution pattern of these spurious sources in terms of the apparent magnitude of the bright star and the distance to it. The~visual analysis led to the definition of a {\it bright star} as an object fulfilling the following two heuristic criteria (Figure \ref{spikes}.), see Solano~et~al. (2022) \cite{Solano2022} for details:
    
    \begin{itemize}
        \item   A source whose magnitude in any of the USNO B,R bands is brighter than \mbox{12.4 magnitudes
        and}
        \item A source whose brightest magnitude in the USNO B,R bands fulfills that
\begin{equation}
            m \leq -0.09* \delta+15.3
        \end{equation}
        
        where {\rm m} is the brightest USNO magnitude (magnitudes) and $\delta$ is the separation (arcsec) between the bright star and the USNO~source.
       
    \end{itemize}
     These two conditions are conservative enough to ensure (at the cost of having some degree of contamination) that real sources are not removed;
    \item Removal of USNO sources not associated with POSS sources: We built a catalog of POSS sources by running {\it Sextractor} \cite{Sextractor}. To~keep faint sources, a~low threshold (just 2$\sigma$ above the background level) was adopted. In addition, to~minimize the number of artifacts, we demand a maximum separation between the USNO and the Sextractor-POSS source of 3.5 arcsec as well as a signal-to-noise ratio $\geq$10 for the Sextractor \mbox{POSS sources.}
\end{itemize}

After applying these filters we ended up with 68\,632 sources (45\% of the original sample), ready to be analyzed by~citizens. 

\unskip
\section{The Web~Interface}\label{sec:interface}

The citizen science project is accessed through the VASCO web interface \cite{Pelckmans2021}. The VASCO web interface differs from the usual {\it Zooniverse} web interface in that each ``mission'' takes longer to fulfill and has more steps. A~bigger emphasis has also been put on the playability aspect of the interface to enhance the entertainment factor~\cite{Pelckmans2021}.

\begin{figure}[H]
   \resizebox{\hsize}{!}{
  \includegraphics[scale=0.3]{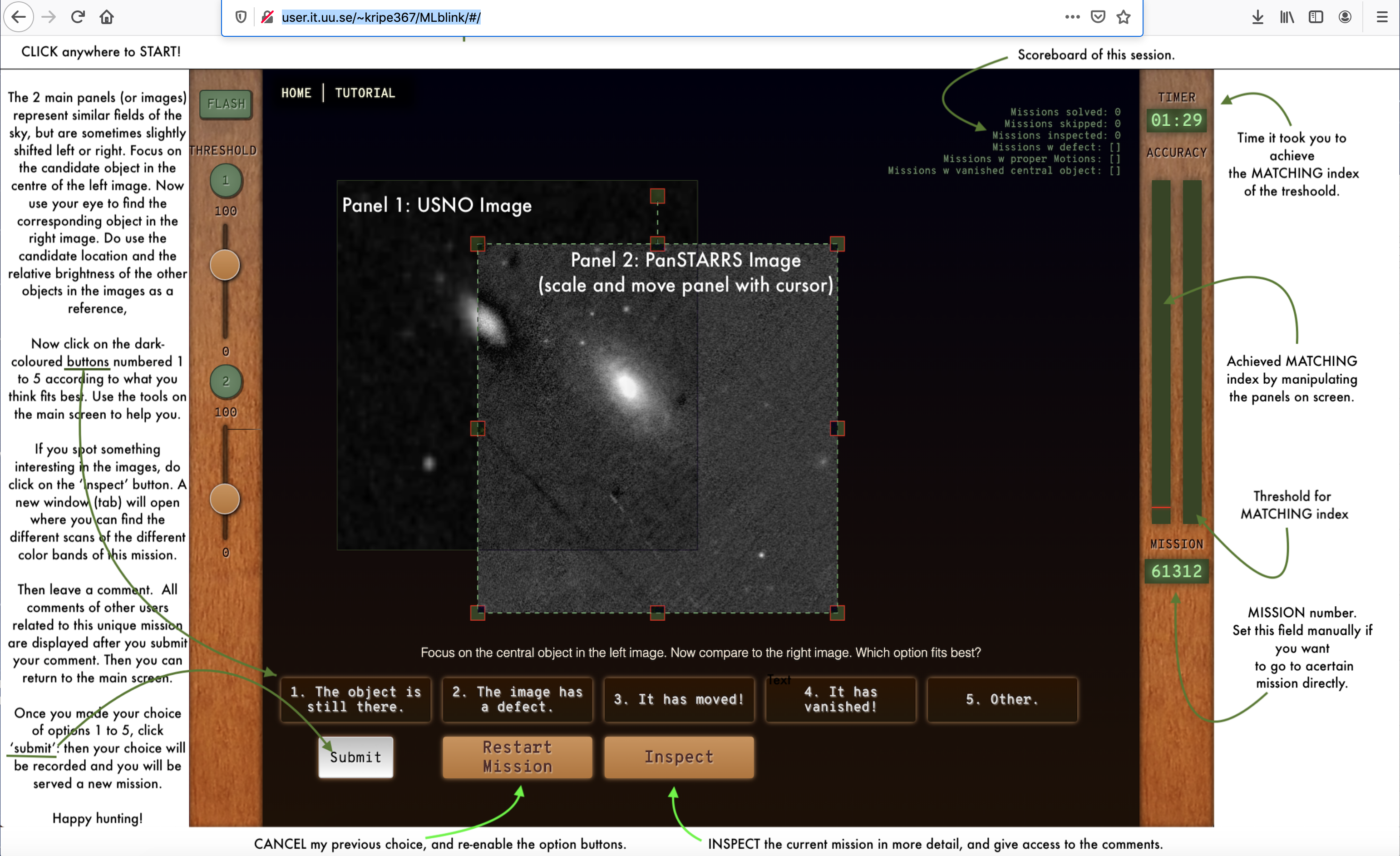}}
  \caption{\label{Splashscreen} The splashscreen from the citizen science project's~webpage.}
   \end{figure}

\begin{figure}[H]
   \resizebox{\hsize}{!}{
  \includegraphics[scale=0.32]{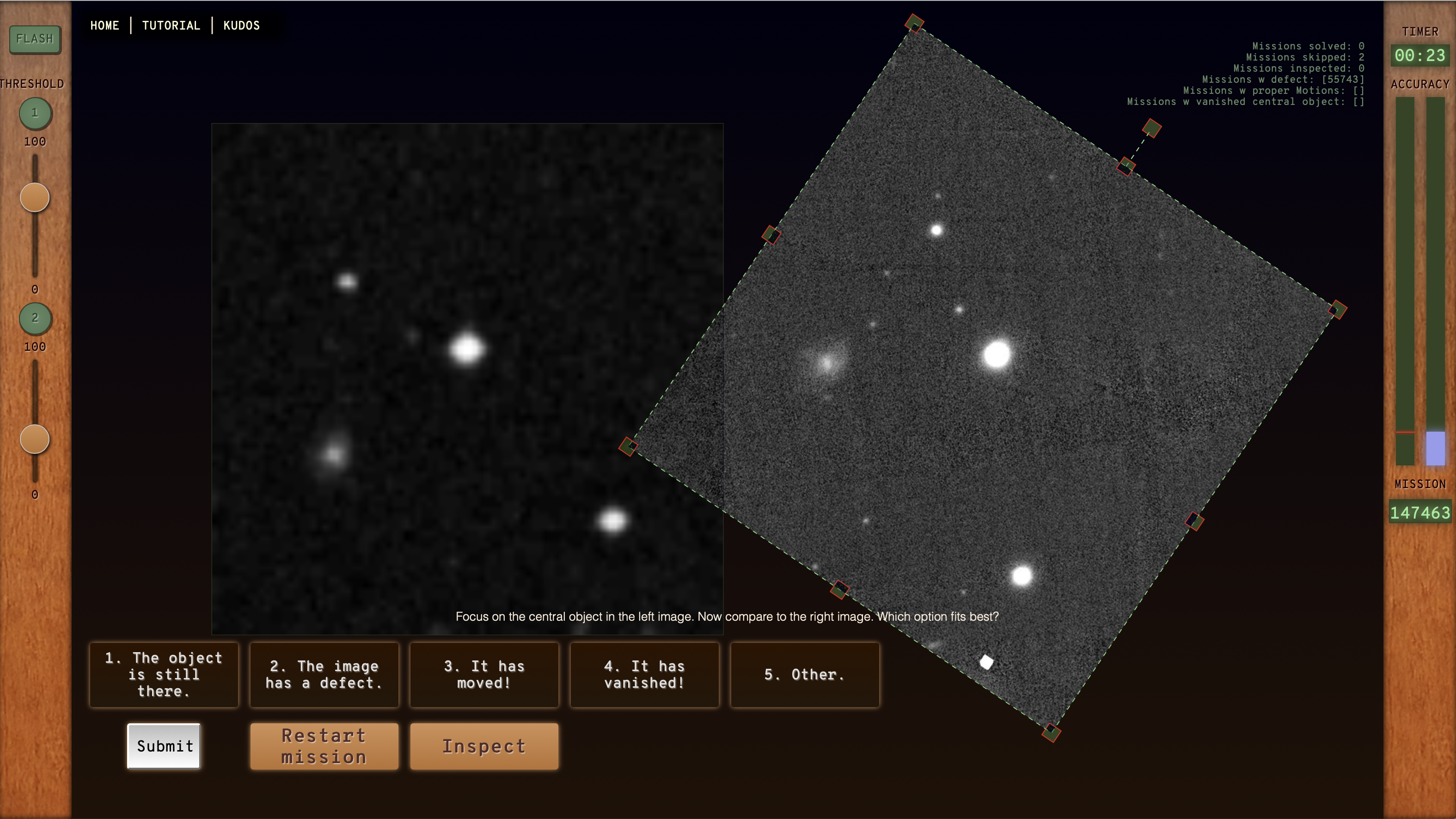}}
  \caption{\label{Rotated} The main active window. Here, we see two images that have been slightly rotated relative to each other. The~user can change the size of the right image, rotate it, and turn it around. The~user can also blink the images by pressing the ``FLASH'' button in the upper left part of the~screen.}
   \end{figure}

A brief guide upon arrival to the web page is immediately given through a splash screen (Figure \ref{Splashscreen}). Once the user has clicked on the splash screen, he or she can engage in examining the images (Figure \ref{Rotated}).

Each user can decide how deep they wish to study each candidate. The~web interface presents the user with a random pair of images, where the left shows an old image and the right, a~new image. Old blue are compared with modern blue images, and~old red with modern red images. Every time a new pair is shown, the~system randomizes which color band that is displayed. The~old and new images have different photometric depths, which the user is asked to take into consideration. In~the current implementation, the~old images are taken from the POSS surveys and the new images from Pan-STARRS, in~order to study the candidates identified by Villarroel~et~al. (2020) \cite{Villarroel2020}. A~user is asked to investigate the images in several different steps:

\begin{enumerate}
    \item Has the central object POSS image vanished from the Pan-STARRS image? (Mandatory/Easy);
    \item Match the two images. The~user is asked to matched the two images as carefully as possible by rotating, turning and changing the sizes of the images. The~matching can be measured through the left accuracy bar (in green). (Optional/Easy);
    \item Inspect. {The} user can investigate the same field in 10 different color bands from the POSS and Pan-STARRS surveys. Here the user can comment upon any interesting details of the images, beyond~the primary science case of a vanishing star. (Optional/Medium difficulty-difficult).
\end{enumerate}

   \unskip

Each of these steps allow for anomaly detection through hypothesis-driven science, exploratory science, and AI-driven science. Moreover, the~variety of approaches that a user can choose covers the search space widely discussed by SETI papers~\cite{Sheikh2020,Singam}.


An underlying artificial intelligence (AI) aimed to help the selection of the most interesting images for the users in current training. The~artificial intelligence learns from the users' image treatment. The~main principles and theory behind the design and structure of the web interface are outlined in Pelckmans~et~al.~\cite{Pelckmans2021}. The~implementation of the AI into the webpage is described by Castillo~\cite{Castillo2019}.

\begin{figure}[H]
   \resizebox{\hsize}{!}{
  \includegraphics[scale=0.32]{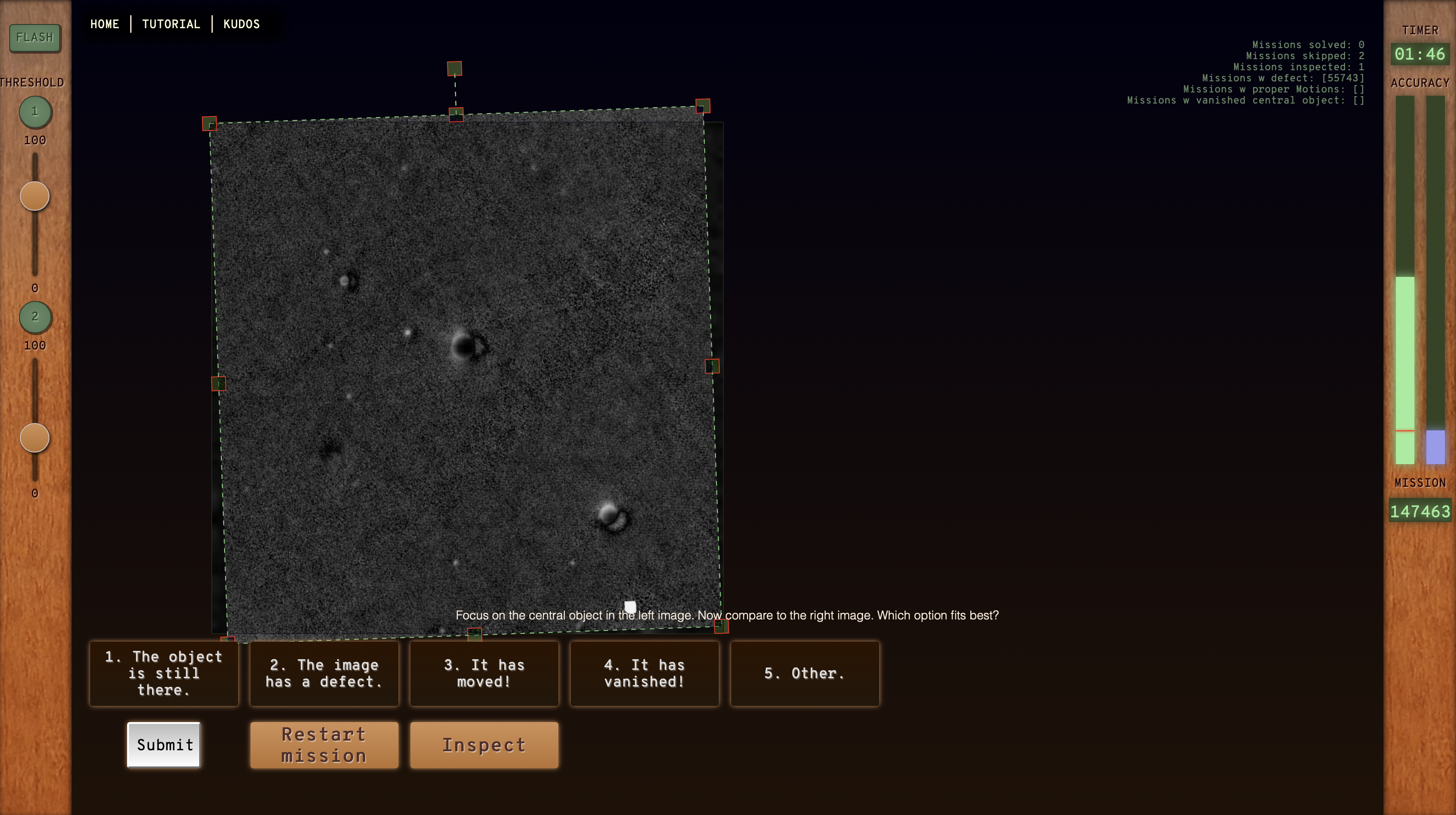}}
  \caption{\label{Matched} An example of two matched images. The~matching of the user results in a high accuracy (green bar), better than that given by the AI (purple bar).}
   \end{figure}

To bring forward the most interesting candidates, the~AI matches the two images and calculates a {\it matching index} that shows how well the two images match in the most central part of the image (see the right ``Accuracy'' bar in the Figure~
\ref{Splashscreen}). As~a comparison, a~user's manual matching is shown in the left accuracy bar next to it. The~user's goal is to obtain a match better than the AI does. Images with the lowest matching index are generally deemed as interesting and worth following up on~\cite{Pelckmans2021}. An~example of a matched pair of images is shown in Figure~\ref{Matched}.

The user can choose from five options:

\begin{enumerate}
    \item ``The object is still there.''
    \item ``The image has a defect.''
    \item ``It has moved!''
    \item ``It has vanished!''
    \item ``Other.''
\end{enumerate}

\begin{figure}[H]
   \resizebox{\hsize}{!}{
  \includegraphics[scale=0.35]{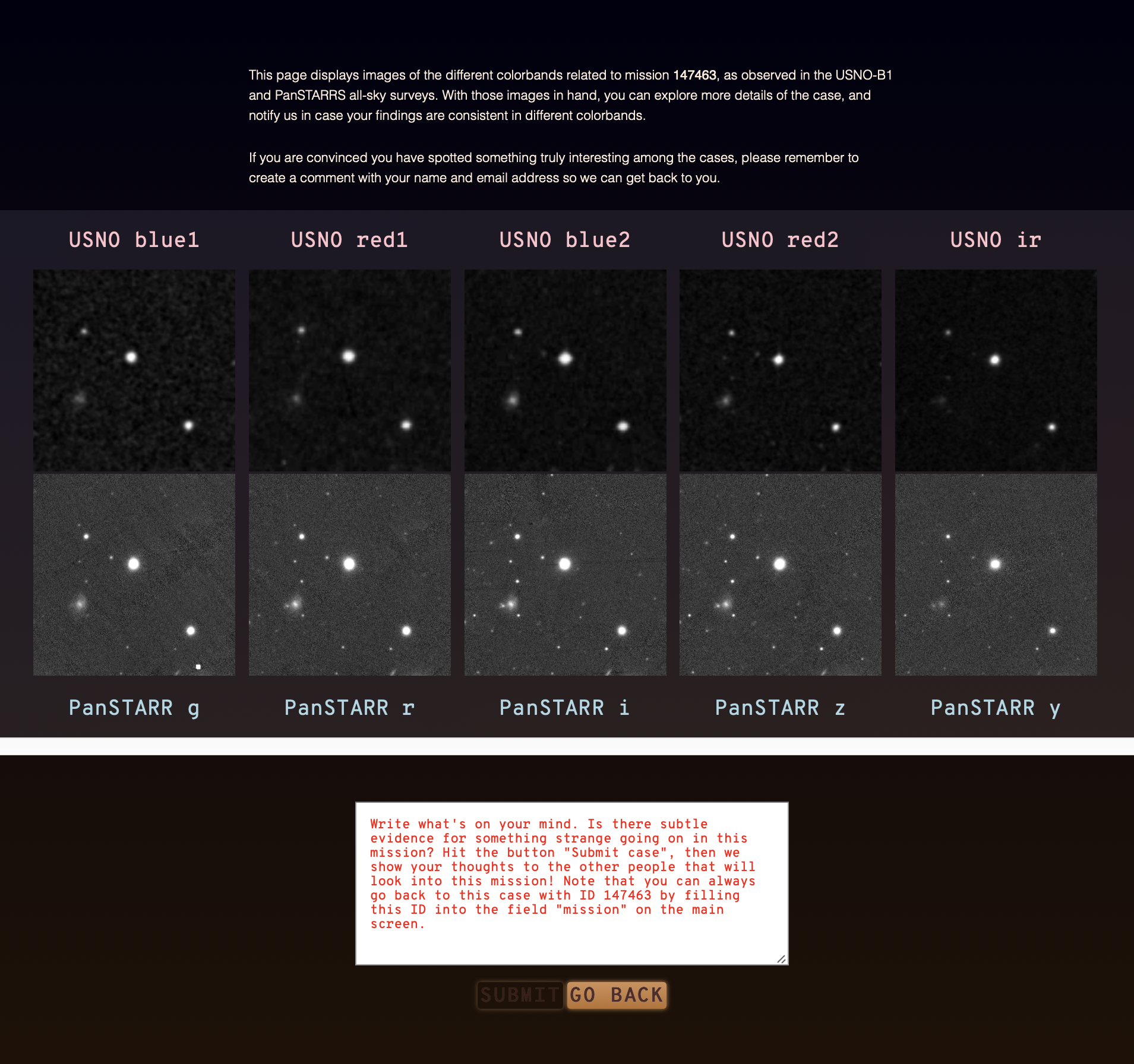}}
  \caption{\label{Inspect} After pressing the ``Inspect'' button, 10 different images from the POSS surveys and Pan-STARRS are shown. The~user is advised to do a careful comparison, take into account the different depth of the given images and asked to remark upon any unexpected findings in the commentary~window.}
   \end{figure}

Sometimes, an~object might appear to have moved, because~the fields-of-view of the images have been rotated. By~rotating the images with the help of the two small squares attached to the Pan-STARRS image, the~user can investigate whether the central star actually moved or if the field of view orientation gave rise to such an effect. Other times, defects might plague one or both of the images. Sometimes, the~user might note something remarkable he or she cannot put words on, in~which we case we ask the user to mark it as ``Other'' and advise them to use the ``Inspect'' button, which opens a new window with a commentary field (Figure \ref{Inspect}). A~tutorial and a tutorial video~\endnote{\url{https://www.youtube.com/watch?v=eM84b6-Z\_xY}}$^{,}$\endnote{\url{https://www.youtube.com/watch?v=gtuF9ISAMRE}} 
is accessible on the webpage. Another tutorial aimed for educational use can be obtained by request.
   
\section{The Citizen Science~Effort}

A citizen science project that generates a large quantity of data also requires a large interactive effort to succeed. The~VASCO citizen science project is public and welcomes all interested users to participate~\endnote{\url{https://www.su.se/english/research/research-news/look-to-the-sky-and-help-researchers-in-a-new-citizen-science-project-1.496340}}. The web page also has a French version~\endnote{\url{http://user.it.uu.se/$\sim$kripe367/MLblink.FR/\#/}}, which makes the project accessible for volunteers in French-speaking countries. We will expand the web interface to offer versions in more languages, e.g., Spanish, over~time.

The mission of the citizen science project can be easily adapted to strikingly different levels of the user's astrophysics background. In~its simplest form, a~user can just play with matching two images and attempt to see if a star has vanished or not, which is a goal achievable for children in their early school years. The~``Inspect'' part of the project, is however having a much more challenging theme, where one has a higher probability of identifying something truly interesting if one has a solid astrophysics background. This step includes images from different epochs and in different color bands, and~may be suitable for university astronomy undergraduate students as a step in training their ability to recognize and identify various type of astronomical sources, but~is not limited only to astronomy~undergraduates. 

\begin{figure}[H]
   \resizebox{\hsize}{!}{
  \includegraphics[scale=0.75]{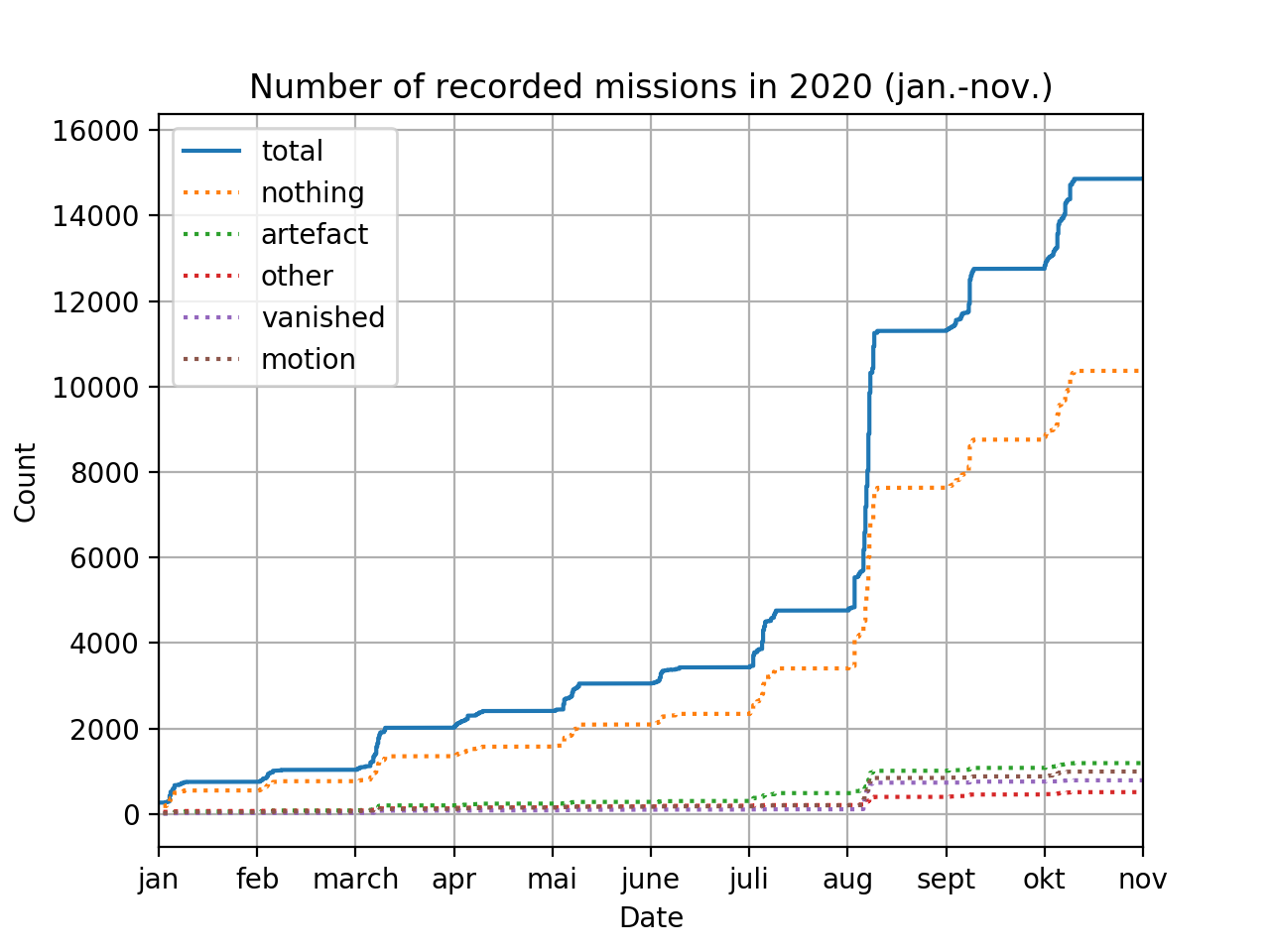}}
  \caption{\label{total} The
 performance over time. We show the number of classifications made by the web interface from the early onset of the project, including the beta testing stage between January and March. The~number of cases labeled as ``vanished'' by the citizens is 798. About 359 additional sources have been highlighted as interesting in ``Inspect''.}
   \end{figure}

The current working mode of VASCO is to collaborate with selected amateur astronomy associations, institutes and educational centers. Figure~\ref{total} shows the number of classifications made by the citizen science project as a function of time, which shows a strong recent increase in the number of classifications. At~the time of writing, we have obtained about 15,593~classifications.

All classifications are counted equally, regardless of who made the discovery. At~the end of the first phase of the project, when classifications have been gathered for all 150,000 objects, the~astronomers in the project will vet each interesting candidate that has emerged in the citizen science project.

\subsection{Collaborating with Schools and Amateur~Associations}\label{sec:Sirius}

A good citizen science project should preferably be a two-way street, where the interactivity is for the benefit of the user as well as for the scientists. A~satisfied user will feel engaged in the project---including the research outcomes---and feel good about his or her contribution to the scientific question. He or she will also learn during the process and feel that there is always more to learn about astronomy and always more fascinating objects to discover. For~the users who are the most engaged, we have included the ``Inspect'' option where the user can investigate the case in 10 different color bands taken at different times. If~a discovery is made, we strongly encourage the user to also submit contact details, so that he or she can be involved in the follow-up studies of the object.
This makes it possible for any person who has made an important discovery to be part of the VASCO research team itself and to be granted credit for the discovery. We also provide user feedback in this fashion. Similarly, scientists will have an increased chance to succeed in the goal of finding anomalies by involving more interested users and thereby examining more~candidates.

While we do not currently have a big platform with registered users, we are in close contact with selected groups of students and amateur astronomers who wish to participate in the VASCO project. This allows us to interact more closely and adapt the~programs.

Collaboration with students can be a good strategy for two reasons; (1) students/pupils can participate as part of their education and are therefore more likely to actually spend the time that is necessary to learn how to make an optimal contribution; (2) they are good control groups for evaluating the effectiveness of the citizen science effort when supervised by their~teachers.

Collaborating with amateur astronomer associations has different advantages. Many amateurs are often engaged in astronomical activities in their free time due to their natural interest. Many often have good observational skills and background knowledge obtained through years as dedicated amateurs. This means that they might be more engaged to start with and therefore produce faster and higher-quality results. Through interaction with amateurs and students, we hope the project also will inspire and engage them to learn more about transient~astronomy.

 \textls[-15]{We are collaborating with student groups and amateur associations in several~countries.}

\subsubsection{Center for Basic Space~Science}
 In Nigeria, the VASCO citizen science project was organized by Center for Basic Space Science (CBSS) with a supporting grant from IAU/Office of Astronomy for Development. The~participants come from Nigeria and Cameroon with a total of 30 people participating including both students and amateur astronomers from different science backgrounds. Each participant conducted the research from his/her home due to the COVID-19 pandemic. The~supporting grant was given to the participants to acquire WiFi, which they used for the project. At~the end of the project about 3000 images were analyzed by the participants. Some of the participants also reported the discovery of some interesting changes between the Pan-STARRS and USNO images in the~process.
 
 \subsubsection{IES~Tartessos}
 
 In Sevilla, a~group of students from the Instituto de Enseñanza Secundaria, IES Tartessos
is participating in the citizen science project since October. The~22 students, about \mbox{12--13 years old,} follow the project with great interest. They report being very thankful for joining this type of collaboration, as~they feel they are helping the scientists leading the research, and~that this has woken up their scientific interest and that they wish to learn more about our~Universe. 

The VASCO sessions are being periodically held once per month during the physics classes. In~order to work, the~students use tablets from the institute. Each participant has to write the number of the ``mission'' in a table, the~date, and~the result of the mission. In~the case that a mission yields an interesting candidate, the~students analyze it once more together with their physics teacher, and~if confirmed once more, the~teacher informs the principal investigators of the project. Until~now, 700 candidates have been treated. By~holding a direct communication, the~students are excited to collaborate in the scientific~project.

\subsubsection{The Sirius Astronomy~Association}
In Constantine, Algeria, a~subset of members from the Sirius Astronomy Association has classified thousands of images. The~members in the Sirius association report that the interest to connect with VASCO is due to its intriguing aspect with non-conventional astronomy and even searches for extraterrestrial intelligence. They also enjoy the multidisciplinary aspect of the project. The~23 strong team from Sirius that is working with VASCO is diverse with ages ranging between 16 and 40, most of them undergraduate students, with~some being PhD students and high school students. The~23 strong group consists of  48\% women and 52\% men. The~members have diverse academic backgrounds, ranging from natural sciences and engineering to humanities and business-related~fields.

For the VASCO citizen science project, the~team is divided into groups which are in turn sub-divided into pairs. This last division is to enable double checking of the results in addition to ensuring that the job is done in case, for whatever reason, one of the members fails to report. The~image sets are then distributed among them, ranging from 10 to 15~images for each pair at the beginning to later reach  25 to 35 images at least. The~members are given a maximum of one week to turn in their work in the form of a text file. The~work of every member is then reviewed by the teacher/leader of the team who makes sure that the images are treated in the proper way. Every image has also been carefully inspected through the ``Inspect'' button. The~members are encouraged to comment whenever needed. Feedback is given to each team member through additional Zoom~meetings.

Training took place through dedicated virtual workshops. These meetings went much beyond the data processing aspect, branching into topics such as the stellar life-cycle, the~formation of nebulae and star clusters, extraterrestrial life and astrobiology, the~basics of spectrometry and its applications to detecting exoplanets and studying organic molecules in the ISM. It also covered the essential principles of AI (artificial intelligence) and its applications to astronomy, in~view of the significant AI component in~VASCO.

More than 1000 images were treated during 2 months of activity, and~the members in the Sirius association are ramping up the pace. They have also implemented a scheme of ``record breaking'' challenges to motivate the team members to treat more images by beating their own~records.

The Sirius organization also has a side activity, ``Learn through VASCO'', where members are encouraged to give talks about topics related to VASCO underpinnings and goals such as the life-cycle of stars, the~Fermi paradox, the~Kardashev scale, Dyson \mbox{spheres, etc.} 

\subsubsection{Vetenskapens Hus and Other~Societies}
Individual students and amateur astronomers from other countries are also participating in the project. The~Soci\'{e}t\'{e} Astronomique de France and Sociedad de Astronomia de Puerto Rico (PRAS) have efficiently involved their members to take part in the citizen science~project. 

In Sweden, we are collaborating with ``Vetenskapens Hus'' (House of Science). Vetenskapens Hus is an educational center in Stockholm. They provide activities for students of all ages; from primary school to high school, as~well as further training for teachers. The~VASCO citizen science project has been included as a part of a course for teachers offered by the center. The~course was a one-day seminar event, which was held for the first time on 11 November 2020. It began with lectures by VASCO members, followed by a hands-on exercise where the participating teachers learn to use the web interface. The~seminar was closed with a discussion session aimed at developing ideas around how the VASCO citizen science project could be implemented in science curricula at various~levels. 

The course at Vetenskapens Hus is designed for a Swedish audience only. However,~the possibility of arranging similar seminar events online and in English, with~a wider and international audience, has been discussed. The~practicalities surrounding such an event have not yet been addressed, however.

\subsection{Social~Media}\label{sec:socialmedia}

In the current time of the Covid-19 pandemic, building ``virtual'' networks is necessary as many schools and universities are closed, and~social distancing is encouraged. Much of the organization of a citizen science effort can be accomplished online. Social media can be exploited to market the project and enable efficient communication. The~VASCO citizen science project has a Facebook page for interested users \endnote{\url{https://www.facebook.com/vascoproject}}.

This is not without challenges, though. The~advertisement of the project through social media relies on scientists being comfortable and experienced users of social media. Another problem is that not everyone has access to reliable internet connections. In~some countries, a~large part of the population may not even have access to a computer and steady WiFi at home. Cell phones seem to be more common, however. In~the future, we hope to be able to adapt the platform to work via~smartphone.

\section{Future Development of the Web~Interface}\label{sec:future}

The VASCO web interface is an easy to use tool that allows users to participate in citizen science projects. Although~more than 15,000 classifications have been made so far, there are still many components of the interface that can be improved. For~instance, there is still no way to compare how each one of our different users have been contributing to our classifications. Additionally, there is no way for users to save their sessions or to edit and update their responses. As~such, in~the future we plan to adapt the interface to support multi-user accounts, allowing users to interact with one another, to~save their sessions, to~upload their profiles into our platform, and~to create their own community of citizen science~friends.

Another natural step is to ``gamify'' (as in user entertainment) the web interface and to create incentives for users to analyze more images. For~example, a~competition feature could allow future users to create internal and public challenges among their community of friends. Together with points, badges, leaderboards, and~performance charts, the~web interface would motivate users to share and evaluate more challenging images. Alternatively, an~invite button would allow new users to join the community through member invitations. All these features would turn the web interface into a modern platform where users participate in the evolution of a community itself, allowing it to expand into directions we cannot think of~today.

\section{Final~Remarks}\label{sec:outlook}

The number of interesting candidates resulting from the new, upcoming cross-match processes may reach millions, placing the project into the regime of ``big data''. VASCO is therefore working to adopt methods from the Virtual Observatory, and~on the further development of an artificial intelligence aided by visual inspection of candidates by citizen~scientists.

The VASCO citizen science project has now launched and together with schools and amateur associations we are orchestrating a community effort to search for anomalies in astronomical images separated by 70 years. It combines both exploratory-driven and hypothesis-driven approaches to the identification of astrophysical anomalies, with~a particular focus on searching for vanishing and appearing~objects.

We plan optical follow-up observations of emerging candidates using, e.g.,~the 3.6m Devasthal Optical Telescope (DOT) at Aryabhatta Research Institute of Observational Sciences (ARIES), Nainital, India. The~aim of the new observations is to examine the nature of the candidates and see if they have faint counterparts ($r > 23.4$ mag) within \mbox{3 arcsec}.

We present the first 15,593 classifications. The~citizen science project will refine the methods for candidate selection and include new data sets with~time.


\vspace{6pt} 


\authorcontributions{B.V. has led the teams' manuscript writing effort and VASCO project design. 
K.P., M.L., A.S., J.S., D.C., S.C. and B.V. developed the citizen science web interface (in order of contribution). E.S. has developed the image pre-processing. J.M., O.N.D. are leading the VASCO
project and citizen science activities in Algeria, Nigeria and Cameroon. K.L. has helped with the coordination of citizen science activities. A.L.G. is leading the citizen science activities at IES Tartessos. L.M. is coordinating the citizen science collaboration with Vetenskapens Hus. S.C., S.G., G.M., A.C.G. are in charge of follow-up observations. R.A. has helped with candidate identification. E.S., G.M., J.M., K.L., O.N.D., A.L.G. and M.E.S. have done substantial efforts in the manuscript writing. All authors assist the citizen science project activities and/or science follow-up. All authors have read and agreed to the published version of the manuscript.}

\funding{This research  made use of the Spanish Virtual Observatory, (\url{http://svo.cab.inta-csic.es}) available online (accessed on October 22 2022). 
 supported from Ministerio de Ciencia e Innovación 
through grant PID2020-112949GB-I00. B.V. is funded by the Swedish Research Council 
(Vetenskapsr\aa det, grant no. 2017-06372) and is also supported by the 
The L’Or\'{e}al---UNESCO For Women in Science Sweden Prize with support of the 
Young Academy of Sweden and the L’Or\'{e}al---UNESCO International Rising Talents prize. 
She is also supported by M\"{a}rta och Erik Holmbergs donation. 
Nordita is partially supported by Nordforsk. M.E.S. acknowledges financial 
support from the Annie Jump Cannon Fellowship, supported by the University of 
Delaware and endowed by the Mount Cuba Astronomical Observatory.
We thank the IAU/OAD for supporting the participation of West Africans in this project, 
through the WAROAD office with the OAD special COVID-19 grant.}

\dataavailability{Not applicable.}

\acknowledgments{We thank Ruben Cubo for help with developing the VASCO web~interface.
This citizen science effort has been driven by many individuals and societies 
who have joined our effort towards finding vanishing stars. We thank PRAS in 
Puerto Rico and Soci\'{e}t\'{e} Astronomique de France (SAF) for helping with 
the project, as~well as all our friends and colleagues who have helped to 
spread information about the~project.
 We thank the Uppsala University Department of Information Technology for support
of the project and VASCO workshop 2018. The~Pan-STARRS1 Surveys (PS1) and the 
PS1 public science archive have been made possible through contributions by the Institute for Astronomy, the~
University of Hawaii, the~Pan-STARRS Project Office, the~Max-Planck Society 
and its participating institutes, the~Max Planck Institute for Astronomy, 
Heidelberg and the Max Planck Institute for Extraterrestrial Physics, Garching, 
The Johns Hopkins University, Durham University, the~University of Edinburgh, 
Queen's University Belfast, the~Harvard-Smithsonian Center for Astrophysics, 
the Las Cumbres Observatory Global Telescope Network Incorporated, the~National 
Central University of Taiwan, the~Space Telescope Science Institute, the~National 
Aeronautics and Space Administration under Grant No. NNX08AR22G issued through the 
Planetary Science Division of the NASA Science Mission Directorate, National Science 
Foundation Grant No. AST-1238877, the~University of Maryland, Eotvos Lorand University (ELTE), 
the Los Alamos National Laboratory, and~the Gordon and Betty Moore Foundation.}

\conflictsofinterest{The authors declare no conflict of interest. The~funders had no role in the design of the study; 
in the collection, analyses, or~interpretation of data; in the writing of the manuscript, 
or in the decision to publish the~results.}

\sampleavailability{Samples of the compounds $\dots$ are available from the authors.}


\abbreviations{Abbreviations}{
The following abbreviations are used in this manuscript:\\

\noindent 
\begin{tabular}{@{}ll}
MDPI & Multidisciplinary Digital Publishing Institute\\
DOAJ & Directory of open access journals\\
TLA & Three letter acronym\\
LD & Linear dichroism
\end{tabular}
}
\begin{adjustwidth}{-\extralength}{0cm}
\printendnotes[custom] 

\newpage
\reftitle{References}

\end{adjustwidth}
\end{document}